\documentclass[10pt,letterpaper,conference]{IEEEtran}

\usepackage{graphicx}
\usepackage{amsmath}
\usepackage{amssymb}
\usepackage{amsxtra}
\usepackage{epsfig}
\usepackage{amsxtra}
\usepackage{amsfonts}
\usepackage{url}

\usepackage[active]{srcltx}
\usepackage{epsf,graphicx}
\usepackage{multicol}
\usepackage{bm}

\newtheorem{thm}{Theorem}
\newtheorem{prop}[thm]{Proposition}

\newtheorem{lem}[thm]{Lemma}

\newcounter{mytempeqncnt}
\begin{document}
\title{On Capacity Computation for the Two-User Binary Multiple-Access Channel}

\author{
\authorblockN{J\"org B\"uhler}
\authorblockA{Heinrich-Hertz-Chair for Mobile Communications \\
Technical University of Berlin\\
Einsteinufer 25, D-10587 Berlin, Germany\\
Email: joerg.buehler@mk.tu-berlin.de}
\and
\authorblockN{Gerhard Wunder}
\authorblockA{Fraunhofer German-Sino Lab \\ for Mobile Communications\\
Einsteinufer 37, D-10587 Berlin, Germany\\
Email: wunder@hhi.fraunhofer.de}
}

\maketitle
\def\thefootnote{\fnsymbol{footnote}}
\footnotetext{This research is supported by Deutsche Forschungsgemeinschaft (DFG) under grant WU 598/1-1.}

\begin{abstract}
This paper deals with the problem of computing the boundary of the capacity region for the memoryless two-user binary-input binary-output multiple-access channel ($(2,2;2)$-MAC), or equivalently, the computation of input probability distributions
maximizing weighted sum-rate. This is equivalent to solving a difficult nonconvex optimization problem. For a restricted class of $(2,2;2)$-MACs and weight vectors, it is shown that, depending on an ordering property of the channel matrix, the optimal solution is located on the boundary, or the objective function has at most one stationary point in the interior of the domain. For this, the problem is reduced to a pseudoconcave one-dimensional optimization and the single-user problem.
\end{abstract}

\section{Introduction}
For some multiuser channel models, the capacity region can be characterized in terms of mutual information expressions. However, even for channels where such a single-letter representation is available, evaluation of the capacity region is often a difficult problem since computation of the capacity region boundary is generally a difficult and nonconvex optimization problem. For the single-user discrete memoryless channel, computation of capacity is a convex problem, and several numerical methods that allow to calculate the capacity within arbitrary precision have been developed, e.g. the Arimoto-Blahut algorithm \cite{arimoto72}\,\cite{blahut72}. For the discrete memoryless MAC, no algorithms for the computation of the capacity region boundary are known. A fundamental step in this direction has been taken in \cite{watanabe96}, where a numerical method for calculating the sum-rate capacity (also called total capacity) of the two-user MAC with binary output has been developed. This was achieved by showing that the calculation of the sum capacity can be reduced to the calculation of the sum capacity for the two-user MAC with binary input and binary output and by giving necessary and sufficient conditions for sum-rate optimality by a partial modification of the Kuhn-Tucker conditions. Unfortunately, further generalizations \cite{watanabe02}-\nocite{watanabe_kamoi04}\cite{watanabe09} of this approach to the most general $(n_1,\ldots,n_m;m)$-MAC (with $m$ users, each with an alphabet of size $n_k$) and consequently the subsequent work in \cite{rezaein04} (which generalizes the Arimoto-Blahut algorithm for the sum capacity computation of the $(n_1,\ldots,n_m;m)$-MAC based on the results in \cite{watanabe02}) is partially incorrect. The work in \cite{CalvoPalomar07} considers the computation of not only the sum capacity, but of the whole capacity region of the two-user discrete MAC. Here, the authors show that the only non-convexity in the problem stems from the requirement of the input probability distributions to be independent, i.e. from the constraint for the probability matrix specifying the joint probability input distribution to be of rank one. They propose an approximate solution to the problem by removing this independence constraint (i.e. relaxation of the rank-one constraint), obtaining an outer bound region to the actual capacity region. By projecting the obtained probability distribution to independent distributions by calculating the marginals, one obtains an inner bound region. Even though the authors present some examples where this approach gives the actual capacity region (i.e. the outer bound region, the inner bound region and the capacity region coincide), the result is often suboptimal, and it is not clarified when the actual capacity region is obtained. 
Consequently, the solution of the capacity computation problem for the discrete memoryless MAC remains an interesting unsolved problem, even for the case of two users and binary alphabets.

{\bf Contributions.}
We prove that for a class of $(2,2;2)$-MACs, the weighted sum objective function has at most one stationary point in the interior of the domain. Beside the fact that this is an interesting structural property which gives valuable insight into the general problem, it can also be employed for numerical solutions of the problem. Since the maximum of the objective function on the boundary can be found by solving the single-user problem, it suffices to search for stationary points in the interior of the domain: As there
is at most one stationary point in the interior, methods such as gradient descent can return a suboptimal
solution only if the global optimum is located on the boundary, which is then found by the boundary search. What is more, we prove the statement by showing that the problem in the interior can be reduced to a pseudoconcave one-dimensional
problem, resulting in an efficient optimization procedure for a specified tolerance of deviation from the optimal point for one of the input parameters. We remark that there is numerical evidence for the conjecture that also for the general $(2,2;2)$-MAC, there is at most one stationary point in the interior of the domain, which we unfortunately could not prove. 

{\bf Organization.}
The paper is organized as follows: 
Section \ref{sec:problemformulation} introduces the problem formulation. In section \ref{sec:twouser2BMAC}, we discuss some general properties of the $(2,2;2)$-MAC. We state the Karush-Kuhn-Tucker (KKT) conditions, discuss some relations to previous work in the literature and reformulate the optimization problem in terms of a one-dimensional and the single-user problem. In section \ref{sec:threeparameterMAC}, we show that for a class of $(2,2;2)$-MACs (the 3-parameter $(2,2;2)$-MAC) and weight vectors with $w_1 \leq w_2$, this one-dimensional problem can in turn be reduced to a pseudoconcave problem, also proving that there is at most one stationary point in the interior of the domain. Imposing a further restriction on the channel transition probabilities, we find a closed-form expression for the solution of the one-dimensional problem. Finally, section \ref{sec:conclusions} concludes the paper.

\section{Problem formulation} \label{sec:problemformulation}
The communication model under study is the discrete and memoryless two-user binary-input binary-output multiple-access channel, termed as $(2,2;2)$-MAC in this paper, which is specified by input alphabets $\mathcal{X}_{1} = \mathcal{X}_{2} = \{1,2\} $, the output alphabet $\mathcal{Y} = \{1,2\}$ and conditional channel transition probabilities $p(y|x_1,x_2)$ for
$y \in \mathcal{Y}, x_i \in \mathcal{X}_i$. Let $Q := \left\{\mathbf{q} = (q_1,q_2)^T \in \mathbb{R}_{+}^{2}: q_1 + q_2  = 1\right\}$. 

It is well-known that the capacity region $\mathcal{C}_{\text{$(2,2;2)$}}$ of the $(2,2;2)$-MAC is given by \cite{ahlswede71}-\nocite{liao72}\cite{tc1991}
\begin{equation} 
\mathcal{C}_{\text{$(2,2;2)$}} = \text{Co}\left(\bigcup_{\mathbf{q}_1,\mathbf{q}_2 \in Q} \mathcal{A}(\mathbf{q}_1,\mathbf{q}_2)\right) 
\end{equation}
where $\mathcal{A}(\mathbf{q}_1,\mathbf{q}_2)$ is the set of all rate pairs $(R_1,R_2)^T \in \mathbb{R}_{+}^2$ that satisfy 
$R_1 \leq I(X_1;Y|X_2), R_2 \leq I(X_2;Y|X_1), R_1 + R_2 \leq I(Y;X_1,X_2).$
Here, $\mathbf{q}_1,\mathbf{q}_2$ specify the input distribution by $\mathbf{Pr}[X_u = s] = q_{us}$, where $q_{us}$ denotes the $s$-th component of $\mathbf{q}_u$, $\text{Co}$ denotes the convex closure operation and $I$ is mutual information.

\begin{figure*}[!t]
\normalsize
\setcounter{mytempeqncnt}{\value{equation}}
\setcounter{equation}{8}
\begin{equation}\label{eq:der1}
\frac{\partial  I(Y;X_1)(p_1,p_2)}{\partial p_1} = 
h_1(p_2) + h_2(p_2)\ln \left( \frac{1}{h_3(p_2) + p_1 h_2(p_2)}-1\right)
\end{equation}
\begin{equation}\label{eq:der2}
\frac{\partial  I(Y;X_2|X_1)(p_1,p_2)}{\partial p_1} = - p_2H(a) + (p_2 - 1)H(b) + p_2H(c) - (p_2 - 1)H(d)
 + H(b + p_2 (a-b)) - H(d + p_2(c-d))
\end{equation}
\setcounter{equation}{\value{mytempeqncnt}}
\hrulefill
\vspace*{4pt}
\end{figure*}

The problem we consider in this work is computing the boundary of the capacity region, or equivalently, since the capacity region is convex, the maximization of the weighted sum-rate in the capacity region for a given weight vector $\mathbf{w} = (w_1,w_2)^T > \mathbf{0}$:
\begin{equation} \label{eq:optproblem1}
\underset{\mathbf{r} \in \mathcal{C}_{\text{$(2,2;2)$}}}{\text{ max}} \mathbf{w}^T \mathbf{r}. 
\end{equation}
Each polyhedron region  $\mathcal{A}(\mathbf{q}_1,\mathbf{q}_2)$ is specified by the corner points 
$ C_{1}(\mathbf{q}_1,\mathbf{q}_2) := (I(Y;X_1), I(Y;X_2|X_1) )^T$ and
$C_{2}(\mathbf{q}_1,\mathbf{q}_2) := (I(Y;X_1|X_2), I(Y;X_2)^T.$
It is easily verified that the weighted sum-rate optimization problem formulated above can be stated in terms of optimization over the region defined by the $C_1,C_2$ points as follows: For $w_1 \leq w_2$, it holds that
\begin{equation}
\underset{\mathbf{r} \in \mathcal{C}_{\text{$(2,2;2)$}}}{\text{max}}\, \mathbf{w}^T \mathbf{r}  = \underset{\mathbf{q}_1,\mathbf{q}_2 \in Q}{\text{max}}\, \mathbf{w}^T C_{1}(\mathbf{q}_1,\mathbf{q}_2) 
\end{equation}
and for $w_1 > w_2$, the optimization can similarly be performed by optimizing over the $C_2$ points.

{\bf Notation and conventions.}
For the transition probabilities of the channel, we write
$a := p(1|1,1), b := p(1|1,2),c := p(1|2,1),d := p(1|2,2)$ and $\Delta_1 := a - b, \Delta_2 := c-d$. We denote the natural logarithm by $\ln$, and express
all entropy and mutual information quantities in nats. The binary entropy function is denoted by $H$. Finally, $D(p||q)$ denotes the Kullback-Leibler divergence between two binary probability  functions defined by $p,q \in [0,1]$. Derivatives on the boundary of closed intervals are to be understood as one-sided derivatives.
We assume without loss of generality (w.l.o.g.) that $0 < w_1 \leq w_2$: For the case $w_1 > w_2$, we can use the fact that $I(Y;X_2)$ and $I(Y;X_2|X_1)$ are obtained from  $I(Y;X_1)$ and $I(Y;X_1|X_2)$ by interchanging the roles of $\mathbf{q}_1$ and $\mathbf{q}_2$ and the roles of $b$ and $c$. 
For $\mathbf{q}_1,\mathbf{q}_2 \in Q,$ we define 
\begin{equation}
\Psi(\mathbf{q}_1,\mathbf{q}_2) := \mathbf{w}^T C_{1}(\mathbf{q}_1,\mathbf{q}_2).
\end{equation}
For channels with $a = b$ and $c = d$, it is $I(Y;X_2|X_1) = 0$ for all $\mathbf{q}_1,\mathbf{q}_2 \in Q$; we exclude this degenerate case from investigation. Similarly, for channels with $a = c$ and $b = d$, $I(Y;X_1) = 0$ for all $\mathbf{q}_1,\mathbf{q}_2 \in Q$, and we also omit
this case. 

{\bf Optimization problem.} In the following, we are thus concerned with the optimization problem
\begin{equation}\label{eq:optproblem}
\underset{\mathbf{q}_1, \mathbf{q}_2 \in Q}{\text{max} }\, \Psi(\mathbf{q}_1, \mathbf{q}_2). \\
\end{equation}
Obviously, for $w_1 = w_2$ the problem (\ref{eq:optproblem}) reduces to the sum capacity problem studied
in \cite{watanabe96}\nocite{watanabe02}\nocite{watanabe_kamoi04}\nocite{watanabe09}-\cite{rezaein04}.

\section{The $(2,2;2)$-MAC}\label{sec:twouser2BMAC}
\subsection{KKT conditions for the $(2,2;2)$-MAC; relation to prior work}
The work in \cite{watanabe96}\nocite{watanabe02}\nocite{watanabe_kamoi04}-\cite{watanabe09} is primarily concerned with proving sufficiency of (modified) Karush-Kuhn-Tucker (KKT) conditions. 
Assuming $a,b,c,d \notin \{0,1 \}$ to ensure differentiability, the KKT conditions corresponding to problem (\ref{eq:optproblem}) can be formulated as 
\begin{eqnarray}\label{eq:KKTConditions}
\frac{\partial \Psi(\mathbf{q}_1, \mathbf{q}_2)}{\partial q_{us}} = \Psi(\mathbf{q}_1, \mathbf{q}_2) - w_u, \,\mbox{if} \, q_{us} > 0,\\ \notag 
\frac{\partial \Psi(\mathbf{q}_1, \mathbf{q}_2)}{\partial q_{us} }\leq \Psi(\mathbf{q}_1, \mathbf{q}_2) - w_u,\,\mbox{if} \, q_{us} = 0
\end{eqnarray}
for $u,s \in \{1,2\}$.
It can easily be checked that for any point satisfying (\ref{eq:KKTConditions}), the linear independence constraint qualification (LICQ) holds, 
so that the KKT conditions given above are a necessary condition for optimality.
Note that these conditions are similar to the expressions given in \cite{gallager68} for the single-user problem and in \cite{watanabe96}\nocite{watanabe02}\nocite{watanabe_kamoi04}-\cite{watanabe09} for the MAC sum-rate capacity. 
Unfortunately, the function $\Psi$ is in general not concave (and not even quasiconcave \cite{boyd2004})
, implying that the KKT conditions in (\ref{eq:KKTConditions}) are not necessarily a sufficient condition for optimality and that solving the optimization problem (\ref{eq:optproblem}) is difficult. 

\begin{figure}
\begin{center}
\includegraphics[scale=0.39]{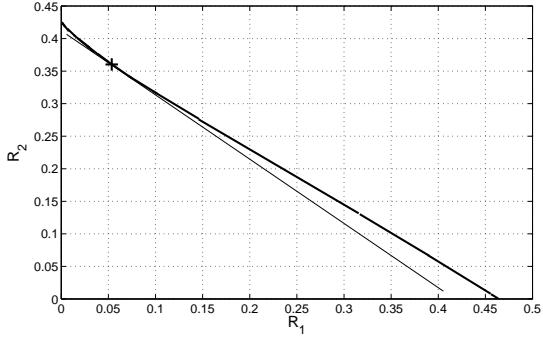}
\caption{Boundary of the nonconvex region $G_1$ for $a=2/3,b=1/4,c=10^{-3},d=5/8$ and $w_1=w_2=1$.}
\label{fig:boundary}
\end{center}
\end{figure}

In \cite{watanabe96}, two classes of $(2,2;2)$-MACs are distinguished: 
case A and case B channels. For case B channels, the KKT conditions as given above (for $w_1 = w_2 = 1$) are proved to be sufficient for optimality. For case A channels, the conditions have to be slightly modified to be sufficient; essentially the modification consists in requiring the optimal point to be located on a certain boundary of the domain. We also note that case A channels are characterized by the condition $(a-c)(b-d) < 0$, and case B channels by $(a-c)(b-d) \geq 0$, although this is not stated explicitly in \cite{watanabe96}. In our case, the situation is quite different: For case A channels, the optimal input distribution is not necessarily located on the boundary. For example, this is the case for the channel with $a=1/5,b=2/5, c=1/2,d=3/10$ and $w_1=1/5,w_2=4/5$. 
The generalization to the $(n_1,\ldots,n_m,m)$-MAC in \cite{watanabe02} and \cite{watanabe09}, where sufficiency of the KKT conditions for elementary MACs (i.e. MACs with $n_k \leq m$ for all $k$) is claimed, is not correct: For example, the $(2,2;2)$-MAC with $a=2/3,b=1/4,c=10^{-3},d=5/8$ and $w_1=w_2=1$ satisfies the KKT conditions in three points, among which actually only one is a global optimum (located on the boundary), one is only a local optimum (also on the boundary) and the third one is a saddlepoint (in the interior). Unlike stated in \cite{watanabe09} (Proposition 3), not every KKT point is a local maximum. Considering the relation to the problem in the rate domain, it is true that every interior KKT point corresponds to a point on the boundary of $G_1 = \{C_1(\mathbf{q}_1,\mathbf{q}_2):\mathbf{q}_1,\mathbf{q}_2 \in Q \}$ (Proposition 2). However, this region is generally not convex unlike implied by the proof of Proposition 3. Figure \ref{fig:boundary}
illustrates this nonconvexity of $G_1$ for the example given above. The marked point on the boundary of $G_1$ corresponds to an interior KKT point and has tangent slope of -1, as indicated by the tangent line drawn in the figure. However, the KKT point corresponding to it is not a local maximum, but a saddle point. 
\subsection{Reduction to a one-dimensional problem}
The formulation of $\Psi$ as a function on $Q \times Q$ served mainly the purpose of relating to prior work.
In the following, we consider $\Psi$ as a function on the domain $[0,1]^2$ instead, i.e. we are concerned with the maximization of $\Psi(p_1,p_2)$ where the input probability distribution is specified by $p_1 = \mathbf{Pr}[X_1 = 1]$ and $p_2 = \mathbf{Pr}[X_2 = 1]$. Our derivation is based on the following observations:
First of all, the boundary points of the capacity region on the two rate axis are $(e_1,0)^T$ and $(0,e_2)^T$, where
\begin{equation} \label{eq:singl1}
e_1 = \underset{i \in \{0,1\}}{\text{max}} \underset{p_1\in [0,1]}{\text{max}}\, I(X_1,X_2;Y)_{p_2 = i}, 
\end{equation} and $e_2$ is given similarly by fixing the value of $p_1$ to 0 and 1. 
Observe that $e_1$ and $e_2$ can be found by solving the single-user capacity maximization problem. 

Furthermore, $I(X_2;Y|X_1)$ is linear in $p_1$, and for (all but at most one, namely the one that satisfies $h_2(p_2)=0$, see below) fixed values of $p_2$, $I(X_1;Y)$ is strictly concave in $p_1$. Hence, the first component in the stationarity equation
\begin{equation}
\label{eq:stationary}
\nabla \Psi(p_1,p_2) \stackrel{!}{=} \mathbf{0}, \, p_1,p_2 \in (0,1)
\end{equation}
has a unique solution in $p_1$ for fixed $p_2$ in the case of strict concavity. What is more, we can find an explicit expression for this solution by simplifying the partial derivative of $I(X_1;Y)$ with respect to $p_1$ such that $p_1$ occurs only once in the expression: The partial derivatives of mutual information at $p_1,p_2 \in (0,1)$ with respect to $p_1$ are given in (\ref{eq:der1}) and (\ref{eq:der2}) at the top of this page. \addtocounter{equation}{2}
Here,
\begin{eqnarray}
h_1(p_2)&:=& H(d + p_2(c-d)) - H(b + p_2(a-b)),\\
h_2(p_2)&:=& -b + d + p_2(-a + b + c - d) ,\\
h_3(p_2)&:=& 1 - d + p_2(d-c).
\end{eqnarray}
We will also write $h_4(p_2) := \frac{\partial  I(Y;X_2|X_1)(p_1,p_2)}{\partial p_1}.$
Let $P_2 := \{p \in (0,1): h_2(p) \neq 0\}$, $\overline{P}_2 := \{p \in P_2: f(p) \in (0,1) \}$.
Note that we have excluded the case $a = c$ and $b = d$, so that there is at most one
$p \in (0,1)$ for which $h_2(p) = 0$. For fixed $p \in P_2$, the explicit solution for $p_1$ in the first component of (\ref{eq:stationary}) is given by $f(p)$, where $f: P_2 \rightarrow \mathbb{R}$ is defined by
\begin{equation} \label{eq:f}
f(p) := \frac{1}{\left(e^{h(p)} +1\right)h_2(p)} - \frac{h_3(p)}{h_2(p)}, 
\end{equation}
with $h(p) := \frac{-\frac{w_2}{w_1} h_4(p)-h_1(p)}{h_2(p)}.$
For $p_2 \in (0,1) \setminus P_2$, it is easy to show that $I(X_1;Y) = 0$ for all $p_1$. Define $\phi: \overline{P}_2 \rightarrow \mathbb{R}$ by $\phi(p):= \Psi(f(p),p)$. Collectively considering the above facts, we obtain
\begin{lem}\label{prop:intermediateopt}
If $\nabla \Psi (p_1,p_2) = \mathbf{0} \mbox{\, for\, } (p_1,p_2) \in (0,1) \times P_2$, then $p_2 \in \overline{P}_2$, $\phi'(p_2)=0$ and there
is no $\tilde{p}_1 \in (0,1)$ such that $\tilde{p}_1 \neq p_1$ and $\nabla \Psi (\tilde{p}_1,p_2) = \mathbf{0}$. Moreover,
\begin{equation}
\underset{p_1, p_2 \in [0,1]^2}{\text{max}}\, \Psi(p_1, p_2) = \text{max} \left\{\underset{p \in \overline{P}_2}{\text{max}}\, \phi(p), w_1 e_1,w_2 e_2\right\}.
\end{equation}
\end{lem}

\section{The 3-parameter $(2,2;2)$-MAC}\label{sec:threeparameterMAC}
The channels that we consider now are $(2,2;2)$-MACs with the restriction $a = p(1|1,1) = p(1|1,2) = b$ on the channel transition probabilities. We call such a channel a {\em 3-parameter $(2,2;2)$-MAC}. The information-theoretic interpretation of such channels
is as follows: Conditioned on the event that user 1 transmits the symbol $1$, the channel that user 2 sees is the single-user antisymmetric binary
channel, which has zero capacity. In other words, whenever user 1 transmits 1, the symbol of user 2 cannot be distinguished at the receiver. In fact, it is easily verified that $I(Y;X_2|X_1=1) = 0.$ If we give the same property to user 1 for user 2 transmitting 1, i.e. $a = b = c$, then $I(Y;X_1|X_2=1) = 0$ and all the points on the boundary of the capacity region can be achieved by "time-sharing between the extremal points on the rate axis". Moreover, we have $e_1=e_2$, so that the capacity region is an isosceles triangle. Furthermore, for $a=b$, we can exchange the values $c$ and $d$ without changing the capacity region. In the following, we thus assume w.l.o.g. that $a \neq c$, $a \neq d$ and $c > d$. 
We also restrict to weight vectors $\mathbf{w} = (w_1,w_2)^T$ with $0 < w_1 \leq w_2$. Unlike in the previous section, this actually is a restriction here: The case $w_1 > w_2$ cannot be treated by exchanging the roles of $c$ and $b$ and $p_1$ and $p_2$, since this would result in a $(2,2;2)$-MAC that is not of 3-parameter type. 

Now consider the extension of $\phi$ to $P_2$, i.e. $\hat{\phi}: P_2 \rightarrow \mathbb{R}$ defined by $\hat{\phi}(p):= \Psi(f(p),p)$ and which is easily verified to be well-defined. 
We will show that $\Psi$ can have at most one stationary point in the interior. We prove this by
showing that $\hat{\phi}$ is pseudoconcave on $(0,1)$. Recall that a twice differentiable function $c: D \subseteq \mathbb{R} \rightarrow \mathbb{R}$ is called {\em pseudoconcave} if $c'(p) = 0 \Rightarrow c''(p) < 0.$ In this case, each local maximum of $c$ is also a global maximum, and $c$ has at most one stationary point. More precisely, we prove 
\begin{prop} \label{thm:intermed}
The function $\hat{\phi}$ has the following properties: 
\begin{itemize}
\item For $a \in (d,c):$ $\hat{\phi}'(p) \neq 0$ for all $p \in P_2$. 
\item For $a \notin (d,c)$ : $\hat{\phi}$ is pseudoconcave on $P_2 = (0,1)$.
\end{itemize}
\end{prop}
\begin{proof}
We first prove the following properties of $h$:
\begin{itemize}
\item For $a \in (d,c): h'(p) \neq 0$ for all $p \in P_2$.
\item For $a \in [0,d): h'(p) = 0 \Rightarrow h''(p) < 0$.
\item For $a \in (c,1]: h'(p) = 0 \Rightarrow h''(p) > 0$.
\end{itemize}
The first derivative of $h$ can be written as
\begin{equation}\label{eq:firstderh}
h'(p) = \frac{\Delta_2 \frac{w_1 - w_2}{w_1}D(a||d+p\Delta_2) + \frac{w_2}{w_1}\delta(a,c,d)}{h_2(p)^2},
\end{equation}
where $\delta(a,c,d):=(c-d)(H(d)-H(a)) - (H(c)-H(d))(d-a)$. For $a \in (0,1)$, 
\begin{equation}
\frac{\partial^2}{\partial a^2}\delta(a,c,d)= (c-d)\left(\frac{1}{1-a} + \frac{1}{a}\right) > 0,
\end{equation}
implying that $\delta$ is strictly convex in $a$. Now $\delta(c,c,d) = \delta(d,c,d) = 0$, so that 
$\delta(a,c,d) > 0$ for $a \notin (d,c)$ and $\delta(a,c,d) < 0$ for $a \in (d,c)$. 
A similar argument shows $\delta(0,c,d) > 0$ and $\delta(1,c,d) > 0$. 
Moreover, $D(a||d+p\Delta_2) \geq 0$ by the nonnegativity of Kullback-Leibler divergence. This implies that 
$h'(p) < 0$ for $a \in (d,c)$. Now consider the situation $a \notin (d,c)$. For $a \in [0,d)$, we have $h_2(p) >0$ for all $p \in (0,1)$ and for $a \in (c,1]$, $h_2(p) < 0$ for all $p \in (0,1)$, so that $P_2 = (0,1)$ for $a \notin (d,c)$. Since
\begin{equation}
h''(p) = \frac{ \frac{w_2 - w_1}{w_1} h_1''(p) - 2\Delta_2 h'(p)}{h_2(p)},
\end{equation}
the claimed property of $h$ follows for $w_1 < w_2$ from the strict concavity of $h_1$. If $w_1 = w_2,$ then $h'(p) > 0$ from (\ref{eq:firstderh}), so that the statement also holds in this case.

We now show that $f(p) < 1$ for all $p \in (0,1)$. For this, we first prove that  $f(p) \neq 1$ for all $p \in (0,1)$. This follows easily for $a \in \{0,1\}$, so that we let $a \in (0,1)$. We also assume $c,d \in (0,1)$; the situations $c \in \{0,1\}$ or $d \in \{0,1\}$ can be treated similarly. It suffices to prove that 
\begin{equation}\label{eq:neg1}
v(p) := \left.\frac{\partial  I(Y;X_1)(p_1,p)}{\partial p_1}\right|_{p_1 = 1} < 0
\end{equation}
and $h_4(p) < 0$ for all $p \in (0,1)$.
To see this, we first note that it can be shown that
\begin{equation}\label{eq:ff1}
v'(p) = \ln\left( \frac{1}{1-a} -1\right) h_2'(p) + h_1'(p) \neq 0
\end{equation}
for all $p \in (0,1)$. 
Moreover, we get $v(0) = -D(d||a) <0$ and $v(1) = -D(c||a) <0$, which together with (\ref{eq:ff1})
imply (\ref{eq:neg1}). For the second statement, observe that
\begin{equation}
h_4''(p) = \frac{\Delta_2^2}{(p \Delta_2 + d)(1-(p \Delta_2 + d))  }> 0,
\end{equation}
so that $h_4(p)$ is strictly convex in $p$. $h_4(p)<0$ then follows from
the fact that $h_4(0) = h_4(1) = 0$. 
Secondly, also using non-negativity of Kullback-Leibler divergence, one can prove 
$\lim\limits_{p \rightarrow 0+}{f(p)} < 1$, which together with $f(p) \neq 1$ and the continuity of $f$ implies that $f(p) < 1$ for all $p \in (0,1)$.

To conclude the proof of the proposition, one can find the following factorized representation for $\hat{\phi}'$:
\begin{equation}\label{eq:factorized}
\hat{\phi}'(p) = w_1(1-f(p))h_2(p) h'(p).
\end{equation}
With the shown properties of $h$ and $f$, the statement follows directly from (\ref{eq:factorized})
for the case $a \in (d,c)$. For the other case, we have that
\begin{eqnarray}
\hat{\phi}''(p) &=& w_1 h'(p)\left(-f'(p)h_2(p)+ \Delta_2(1-f(p))\right)  \\ \notag 
&&{+}\: w_1(1-f(p))h_2(p)h''(p),
\end{eqnarray}
implying that with the properties of $h$ and $f$ follows that
\begin{equation}
\hat{\phi}'(p^{*}) = 0 \Rightarrow h'(p^{*}) = 0 \Rightarrow \hat{\phi}''(p^{*}) < 0.
\end{equation}
\end{proof}
It can be verified that for $h_2(p_2) = 0$, it is $\Psi(p_1,p_2) \neq \mathbf{0}$ for all $p_1 \in (0,1)$. As a consequence, Proposition \ref{thm:intermed} and Lemma \ref{prop:intermediateopt} imply that in the case $a \in (d,c)$ (i.e. for 3-parameter channels of case A), the function $\Psi$ has no stationary point in the interior; the optimum input probability distribution is located on the boundary and it suffices to solve the single-user problems. Speaking in terms of the KKT conditions  (\ref{eq:KKTConditions}), this means that each point satisfying these equations is located on the boundary,
as in the case of the sum-rate problem \cite{watanabe96}.
For the case $a \notin (d,c)$ (case B), pseudoconcavity of $\hat{\phi}$ implies that there is at most one stationary
point (or, equivalently, at most one KKT point) in the interior of the domain. We summarize this in the following theorem:
\begin{thm}
For the 3-parameter $(2,2;2)$-MAC with $a \neq c$, $a \neq d$, $c > d$ and $w_1 \leq w_2$, the following holds:
\begin{itemize}
\item For $a \in (d,c)$, the optimal input distribution is located on the boundary of $[0,1]^2$, and there is no stationary point of $\Psi$ in the interior of $[0,1]^2$.
\item For $a \notin (d,c)$, there is at most one stationary point of $\Psi$ in the interior of $[0,1]^2$.
\end{itemize}
\end{thm}

By Lemma \ref{prop:intermediateopt}, the problem of finding the maximizing input distribution can be reduced to the single-user problem and the optimization of $\phi$ (for which it suffices to optimize $\hat{\phi}$, as described in the following). Since $\hat{\phi}$ is pseudoconcave, it can efficiently be optimized using a simple standard bisection algorithm: For a given tolerance $\varepsilon$, we start with the interval $[\epsilon,1-\epsilon]$ and determine if one of the intervals $[\epsilon,1/2]$, $[1/2,1-\epsilon]$ contains the optimal point by checking the sign
of $h'$ (i.e. the sign of $\hat{\phi}'$ by (\ref{eq:factorized})) at the interval boundaries. 
If this is not the case, we assume the optimal point to be on the boundary. Otherwise, we continue bisecting the interval that contains the stationary point until 
the interval length is smaller than $\varepsilon$, and find a solution $p_{\epsilon}$ within $\epsilon$ deviation tolerance from the optimal using only $O(\log(1/\epsilon))$ evaluations of $h'$, which is much more efficient than a brute-force search. 
By the definition of $\phi$, $(p_1^*,p_2^*)= (f(p_{\epsilon}),p_{\epsilon})$ is used as optimization output if $f(p_{\epsilon}) \in (0,1)$. If $f(p_{\epsilon}) \notin (0,1)$, we assume that the stationary point of $\hat{\phi}$ is outside of $\overline{P}_2$, and we also assume the optimal point for $\Psi$ to be on the boundary. Note that in principle, we could find the optimal $p_1$ for this choice of $p_2=p_{\varepsilon}$ by solving the single-user problem for fixed $p_2=p_{\varepsilon}$. However, it is clear that for sufficiently small $\epsilon$, we will have $f(p_{\epsilon}) \in (0,1)$ if $\Psi$ has a stationary point in the interior.  Note that we can only give a deviation tolerance for $p_2^*$. 
However, $p_1^* = f(p_{\epsilon})$ is still a reasonable solution since it is optimal "conditioned" on the choice of $p_2^* = p_{\epsilon}$. We remark that a gradient descent algorithm employed for $\Psi$ also typically shows fast convergence to the optimal input distribution; we also refer to the introductory discussion.  

We finally consider the 3-parameter $(2,2;2)$-MAC with $a = b = 0, 0 < d < c$ and $w_1 < w_2$.
For this channel, it is possible to give an explicit solution for the optimization of $\phi$.
It can be shown that $\overline{P}_2 = (0,1)$, so that $\hat{\phi} \equiv \phi$.
Here, we can find the zero of (\ref{eq:factorized}) by solving $h'(p)=0$ for $p$, resulting in
\begin{equation}
p = p^{*}(c,d,\mathbf{w}):=\frac{1 - d - e^{-\frac{w_2\delta(0,c,d)}{\Delta_2(w_2 - w_1)}}}{\Delta_2},
\end{equation} 
where
$\delta(0,c,d)=(c-d)H(d) - d(H(c)-H(d)).$
The maximum weighted sum-rate is given by
\begin{equation}
 \underset{\mathbf{r} \in \mathcal{C}_{\text{$(2,2;2)$}}}{\text{ max}} \mathbf{w}^T \mathbf{r}= \left\{\begin{matrix}
\phi(p^{*}(c,d,\mathbf{w})), &  p^{*}(c,d,\mathbf{w}) \in (0,1)\\
\text{max}\{w_1 e_1,w_2 e_2 \}, & \mbox{otherwise.}
\end{matrix}\right.
\end{equation}
\enlargethispage{-1.2mm}
\section{Conclusions} \label{sec:conclusions}
In this work, we studied the problem of maximizing weighted sum-rate (or computing the boundary of the capacity region) for the memoryless two-user binary-input binary-output multiple-access channel, called $(2,2;2)$-MAC. The KKT conditions were formulated as a necessary optimality condition. However, the objective function is not concave, and the KKT conditions are not sufficient for optimality. We demonstrated that this is the case even for the sum-rate problem, unlike stated in prior work. In this paper, we proved some structural properties of the problem. For this, it was first reduced to a one-dimensional problem and the single user problem and then studied for the 3-parameter $(2,2;2)$-MAC, for which $p(1|1,1) = p(1|1,2)$. For weights satisfying $w_1 \leq w_2$, we showed that, depending on an ordering property of the channel transition probability matrix, either the maximum is attained on the boundary, or there is at most one stationary point in the interior of the optimization domain. The proof was obtained by showing that in the latter case, the reduction to the one-dimensional problem leads to a pseudoconcave formulation, which can also be used numerically for the capacity optimization. For a further restricted class of $(2,2;2)$-MAC channels, an explicit solution for the one-dimensional problem could be given. Future work will consider the unrestricted $(2,2;2)$-MAC and the most general situation of the $(n_1,\ldots,n_m;m)$-MAC.

\bibliographystyle{/home/buehler/TeX/BibTex/IEEEtran}
\bibliography{/home/buehler/TeX/BibTex/references}

\end{document}